# Raman scattering as a probe of intermediate phases in glassy networks

**P. Boolchand, Mingji Jin, D.I. Novita, S. Chakravarty**
Department of Electrical and Computer Engineering and Computer Science
University of Cincinnati, Cincinnati, OH 45221-0030, USA

**Bulk glass formation occurs over a very small part of phase space, and "good" glasses (which form even at low quench rates ~ 10K/sec) select an even smaller part of that accessible phase space. An axiomatic theory provides the physical basis of glass formation, and identifies these sweet spots of glass formation with existence of rigid but stress-free networks for which experimental evidence is rapidly emerging. Recently, theory and experiment have come together to show that these sweet spots of glass formation occur over a range of chemical compositions identified as Intermediate Phases. These ranges appear to be controlled by elements of local and medium range molecular structures that form isostatically rigid networks. Intermediate Phase glasses possess non-hysteretic glass transitions ($T_g$s) that do not age much. Raman scattering has played a pivotal role in elucidating molecular structure of glasses in general, and in identifying domains of Intermediate Phases. Experiments reveal these phases to possess sharp phase boundaries and to be characterized by an optical elasticity that varies with network mean coordination number, *r*, as power-law. In this review, we provide examples in chalcogenide and oxide glass systems where these phases along with optical elasticity power-laws have been established. Intermediate Phase glasses represent self-organized nanostructured functional materials optimized by nature.**

**KEYWORDS**: Self-organization in glasses, percolation of rigidity; reversibility window; intermediate phases, Raman scattering, photocontraction.

## INTRODUCTION

Most liquids upon cooling undergo a first order transition at the melting point $T_m$ to a crystalline state. On the other hand, select liquids can be supercooled to bypass crystallization and form a glass at a lower temperature called the glass transition temperature $T_g \sim 2/3$ of $T_m$ [1]. What is so special about these select liquids that readily form glasses? They usually consist of polymeric networks with a mean coordination number, *r* ~2.40. The quantity *r*, is formally defined as $(1/N)\Sigma n_i r_i$, where $n_i$ represents the number of atoms having a coordination of $r_i$, and $N = \Sigma n_i$ gives the total number of atoms in a network. The optimum value of *r* = 2.40, was first introduced from an axiomatic theory of glass formation[2] and has found much support in covalently bonded systems[3]. Experiments reveal that space filling, rigid but unstressed networks do indeed occur near *r*~ 2.40. In the past decade, the recognition has emerged[4-9] that instead of there being a solitary chemical composition[2], there is a range of chemical compositions across which the optimal requirement is fulfilled in real glasses. In this special range, also called *Intermediate Phases (IPs)*, not only is the glass forming tendency optimized but also local and intermediate range structures give rise to networks that are isostatically rigid. Window glass[10], thin-film gate dielectrics[11], thermally reversible folding pathways of proteins[12] include some examples of disordered systems where IPs are manifested[13]. The starting point to understand IPs is to examine the elastic response of glassy networks.

**INTERMEDIATE PHASES IN GLASSY NETWORKS**

In 1788, J. L. Lagrange laid the foundations[14] of generalized coordinates and constraints in Mechanics in the celebrated book "Mechanique Analytique". A century later, J.C.Maxwell showed[15] how these ideas on constraints can be used to understand mechanical stability of macroscopic structures such as trusses and bridges. More recently, and quite independently, ideas on mechanical constraints were extended to disordered covalently bonded atomic networks by J.C. Phillips[2] to understand their glass forming tendency. He suggested, that the glass forming tendency is optimized in structures that are optimally constrained, i.e., when

$$n_c = 3 \qquad (1)$$

Here $n_c$ represents the mean count of bond-stretching and bond-bending constraints per atom, network becomes ideally connected in the sense mentioned in the Introduction when the count equals 3, the degrees of freedom per atom in 3d. The axiom (1) selects structures that are neither highly nor weakly cross-linked, and leads to the magic connectivity of $r = 2.40$ in percolating networks with no dead-ends. M.F. Thorpe showed[16] that axiom (1) actually represents a condition for a floppy network to become mechanically rigid. A normal mode analysis of a weakly cross-linked chain network shows cyclical (or floppy) modes (zeros of the dynamical matrix) to number on average, $f = 3 - n_c$. The average coordination number $r$ (or constraints $n_c$) per atom are known as mean-field values. In a network of long chains, such as found in a Se glass for example, every atom has two nearest neighbors ($r = 2$), and there is one bond-stretching and one bond-bending constraint per atom. This gives $n_c = 2$ constraints/atom; so there is one floppy mode per atom, $f = 1$.

In other words, 1/3rd of the vibrational modes are floppy. Inelastic neutron scattering measurements[17] show a mode near 5 cm$^{-1}$, which carries one-third of the spectral weight in a Se glass. The presence of floppy modes renders this inorganic polymer network to be rather flexible. Cross-linking the chains of Se by alloying either a group IV and/or a group V additive steadily lowers, the floppy mode fraction and, within a mean-field description[2,16], the count vanishes linearly as $r$ increases to a critical value $r_c$ = 2.40. This defines the elastic phase boundary between a flexible and a rigid network. The mean-field prediction on percolation of rigidity near $r$ = 2.40 has been refined in numerical experiments on simple random networks[18] constrained by bond-stretching and bond-bending forces. At $r$ > 2.40, one has a stressed-rigid network in which these numerical experiments[18,19] show shear, longitudinal and transverse elastic constants (C) to increase as a power-law in $r$, i.e.,

$$C - C_o = K(r - r_c)^p \qquad (2)$$

with a power p close[18,19] to 1.50. Here $C_o$ represents the value of C at $r = r_c$. More recently, numerical experiments have been extended to self-organizing networks of carbon or silicon[20]. They show the presence of *two* elastic phase transitions, a *rigidity*- transition ($r_1$), followed by a *stress*-transition ($r_2$). In the intervening region, $r_1 < r < r_2$, the network is rigid but unstressed. This region is an *Intermediate Phase* (IP) distinguished from the *Flexible Phase* at $r < r_1$ and the *Stressed-Rigid Phase* at $r > r_2$. . In the IP, inserting a new cross-link finds it way to the flexible part of a network, thus self-organizing to avoid forming redundant bonds. The avoidance[20] of redundant bonds by reconnecting the network is no longer possible once $r > r_2$. The absence of redundant bonds in the IP ($r_1 < r < r_2$) lowers the free energy, and is the driving force for disordered networks to self-organize. However, the calculated IP widths in amorphous carbon or silicon are much too narrow ($\Delta r$ = 0.016) [21], and the IP centroids always reside at values of $r$ slightly less than 2.40. However, IP widths, close to experimental ($\Delta r$ = 0.12) values[22], have been found in an analytic calculation of realistic alloy structures using a procedure called Size Increasing Cluster Agglomeration (SICA)[23].

A point of current interest relates to aspects of local and intermediate range structural order that control widths and centroids of IPs in glasses. Rigidity in glassy networks can be traced to rings that have typically 4 to 6 atoms[21, 23]. Larger rings tend to be flexible while smaller ones stressed-rigid. The relevant scale of molecular structures is thus typically 10A in size. However, the way rings connect to each other forming a network of super-structure of rings is also relevant. IPs are non-mean-field phases of glassy networks in that more than the closest environment of individual atoms is relevant; and their detection in experiments requires methods that, in general, probe a network at all length scales. Recently, IPs have been found in numerical simulations on triangular networks[24, 25], and the suggestion has been made[25] that IPs may be manifestations of self-organized criticality.

**EXPERIMENTAL PROBES OF INTERMEDIATE PHASES**

The breakdown of the k = 0 selection rule[26] in disordered systems, permits Raman scattering[27, 28] to be a particularly fertile probe of *local* ( ~ 3A) and intermediate range (~ 10A) structures of glassy networks. In some cases these experiments have been an avenue to establish the *three* elastic phases in glasses as discussed earlier[4-7]. When a vibrational mode forms part of a network and has some extended character, one can experimentally define optical elastic power-laws like in equation 2 from variations in mode frequency (ν) as a function of network connectivity *r*.

$$\nu^2 - \nu_c^2 = K\,(r - r_c)^p \qquad (3)$$

Here $\nu_c$ represents the value of ν at the elastic threshold $r_c$. Several binary and ternary covalent glass systems have been examined in this respect, and the results show[4-7] the elastic-power-law in the Intermediate and Stressed-rigid phases to be p ~1.0 and ~ 1.5 respectively. Thus, Raman scattering through a measurement of optic modes as a function of network connectivity has permitted probing glassy networks not only at a local scale, but also at intermediate and even extended length scales. We shall return to discuss this point in conjunction with experimental results later.

When a crystalline solid is squeezed by application of a hydrostatic pressure P, bonds reduce in length and vibrational modes, in general, blue-shift on account of anharmonicity of bonds[29]. In glasses, parallel experiments have demonstrated that vibrational modes usually broaden and also, blue-shift, but only when the external pressure (P) exceeds a threshold value ($P_c$). $P_c$ is generally regarded[30] as a measure of network internal stress in a glass, and the external pressure P must exceed $P_c$ for a glassy network to *feel* the applied pressure. Recent Raman pressure experiments on the binary $Ge_x Se_{1-x}$ glasses have shown [22] trends in threshold pressures, $P_c(x)$; which vanishes in IP compositions, $0.20 < x < 0.25$ and increases steadily once $x > 0.25$ in the stressed-rigid phase or when $x < 0.20$ in the flexible phase. Thus, Raman scattering through pressure measurements using Diamond Anvil Cells has also served as a useful probe of network stress in glasses in general, and IPs in particular.

Brillouin scattering is a powerful probe of acoustic excitations in solids[31] including glasses. In such experiments the intrinsic length scale to which a glassy network is probed is set by the wavelength of a longitudinal or a transverse acoustic excitation, which typically lies in the range of 100 nm or larger in oxide and chalcogenide glasses[32]. Since this length scale exceeds the relevant length scale of rings (~ 10A) structural order in IPs, Brillouin scattering largely serves as a mean-field probe of elastic behavior in glasses[32]. On the other hand, Raman scattering through optic modes can probe glass structure at both short and extended range to detect IPs. For these reasons, IPs are manifested in Raman scattering[8] but not in Brillouin scattering[32, 33] experiments.

A particularly rewarding probe of elastic phases in glasses is modulated Differential Scanning Calorimetry (m-DSC)[34-36]. The thermal probe permits a detailed examination of a glass transition endotherm using in essence AC calorimetry[34]. In AC calorimetry one applies a small sinusoidal heat flux and measures the temperature response at the same frequency, permitting the frequency dependence of the specific heat to be established. In m-DSC, one superposes a sinusoidal temperature variation over a linear temperature ramp to thermally scan across a glass transition, and deconvolutes the *total* heat flow endotherm near a glass transition into two contributions[35,36]; one that tracks the sinusoidal variations and is called the *reversing* heat flow, and the difference signal between the *total* and *reversing* heat flow that does not track the temperature oscillations, and it is reckoned as the *non-reversing* component of heat flow. The reversing heat flow usually shows a step typically 20 to 40 °C wide, and one defines the glass transition temperature ($T_g$) by the inflexion point of the step. The step height gives the specific heat change across the glass transition. The much higher sensitivity of the AC (m-DSC) over the traditional DC measurements (DSC) has the advantage that one can use much lower scan rates in the former (2°C/min) compared to the latter (20 °C/min). The m-DSC method permits obtaining scan-rate and thermal history independent $T_g$s. The non-reversing heat flow usually shows a Gaussian-like profile as a precursor to the glass transition, and the frequency corrected[35,36] area under the Gaussian profile provides a quantitative measure of the non-reversing enthalpy ($\Delta H_{nr}$) of the glass transition. Experiments on a variety of glasses have revealed compositional windows across which the $\Delta H_{nr}$ term nearly vanishes[4,22], and glass transitions become thermally reversing (reversibility window) in character[36]. Glass systems examined in both Raman scattering and m-DSC experiments reveal the rather spectacular result that the IPs inferred in Raman scattering virtually coincide with thermal reversibility windows. In retrospect this is not surprising given the fact that both the thermal and optical method probe glassy networks at all length scales. The correlation between the thermal and the optical probe of IPs has served as an independent check of one method over the other, and together both techniques have provided new insights into the elusive nature of glass transitions[4,37] and self-organization[13] in disordered systems.

**INTERMEDIATE PHASES IN CHALCOGENIDE GLASSES**

Binary and ternary alloys of the group IV (Si, Ge, Sn) and group V (P,As,Sb) with the group VI (S,Se and Te) elements (chalcogens) represent some of the best glass formers in nature. These chalcogenide alloy glasses are easily synthesized by reacting the pure elements. Starting from a pure Se glass, the addition of a group IV and/or a group V element leads to a progressive crosslink of the Selenium chain network to increase its mean coordination number $r$. The behavior is reflected in compositional trends of $T_g$ ($r$) for which theoretical support has emerged from stochastic agglomeration theory[38].

Consider the cases of binary $Ge_xSe_{1-x}$, $P_xSe_{1-x}$ and ternary $Ge_yP_ySe_{1-2y}$ glasses for which compositional trends[39] in $T_g$ appear in Fig.1. One observes thresholds in $T_g$ for the case of the two binary glass systems near their chemical thresholds ($r_t$ = 2.67 and 2.40), but the absence of such a threshold in the ternary glass system near the chemical threshold ($r_t$ = 2.55). The chemical threshold represents the glass chemical composition where available Se is all used up in forming the Ge- and P-centered local structures, thus leaving no excess Se in the network to form $Se_n$-chain fragments. In the ternary glass system, the chemical threshold occurs[6] at $x_t$ = 18.18% corresponding to $r_t$= 2.55. In binary glasses, once $r$ approaches $r_t$, in many instances the cation-cation bonds formed are found to intrinsically segregate from the backbone leading to nano-scale phase separation (NSPS) of glasses. Specifically, in binary $Ge_xSe_{1-x}$ glasses, Ge-Ge bonds and in $P_xSe_{1-x}$ glasses, P-P bonds, apparently demix from the backbone to form nano-phases at $r > r_t$. The loss in network connectivity on account of NSPS in glasses leads to global maxima in $T_g$ near chemical thresholds[40], and is reminiscent of a parallel behavior of the liquidus of corresponding crystals in the equilibrium phase diagrams. These NSPS effects are qualitatively absent[39] in the ternary glass system, demonstrating that Ge-Ge bonds and P-P bonds formed at $r > r_t$ largely form as part of the backbone. In a similar fashion the case of $Ge_xAs_xSe_{1-2x}$ ternary has been extensively investigated[6,8]. Ternary glasses are particularly attractive systems to examine connectivity related phase transitions because of the absence of NSPS effects[39].

The richness of the Raman scattering results in ternary glasses[41] is displayed in Fig.2, which gives the observed lineshapes at low, intermediate, high and very high frequencies in 4 separate panels. The spectra are stacked bottom up with increasing content of P or Ge, and mode identification is provided in Table 1 and the labels in figure 2. These spectra provide a pictorial view of glass molecular structure evolution with x, and the principal features are as follows. At x = 0, one starts with a $Se_n$ chain glass with a finite concentration of $Se_8$ rings[42]. Addition of Ge and P ( x = 3%) leads to the formation of Ge-centered Corner-Sharing (CS) and Edge-Sharing (ES) tetrahedra[43, 44], and P-centered pyramidal (PYR) and quasi-tetrahedral (QT) units[45] that serve to crosslink the chains of $Se_n$. Polymeric ethylene-like $P_2Se_4$ units containing P-P bonds are first manifested[45] near x ~9% and grow rapidly with x. Near x ~20% $P_4Se_3$ molecules are formed[45, 46], which decouple from the backbone in the 18% < x < 24% range nanoscale phase separating (NSPS) from the backbone. At higher x, a variety of new structures form, including ethanelike $Ge_2Se_6$, amorphous Ge[47], amorphous phosphorus[48] and $P_4$ monomers[49, 50].

The feature of principal interest in the Raman spectra of Fig.2 is the CS mode of GeSe$_4$ tetrahedra[43] near 200 cm$^{-1}$. The mode frequency is found to steadily blue-shift as a function of x, a behavior that is mapped in Fig.2(b) and Fig. 3(b). At low x ( <5%) and in region I, glasses age ever so slowly and the mode frequency steadily blue-shifts in time over a period of months, a feature that is characteristic of flexible glasses. But as x increases to 7%, such aging is qualitatively suppressed, and variations in mode frequency, $\nu_{cs}(x)$, display a mild kink (change in slope) near x = 9%. The $\nu_{cs}(x)$ variation is almost linear in the 9% < x < 14.5% range, ( Region II in Fig. 3b). This is followed by a small jump in frequency of about 0.4 cm$^{-1}$ near x = 14%, and a power-law variation in region III, 14.55 < x < 17%. A mild NSPS of the ternary glass initiates at x > 18% as P$_4$Se$_3$ molecules decouple from the backbone[45, 46] and lower the global connectivity (see Fig.3a), interrupting the power-law variation, an issue we discuss next.

The nature of the CS mode of a GeSe$_4$ tetrahedral unit in GeSe$_2$ glass has been examined in ab-initio MD simulations[51], and inverse participation ratios reveal that the mode has some extended character. This is not unexpected given that the Se bridging bond angle across two tetrahedral units deviates from 90°. The structural feature insures that the motions of adjacent tetrahedra are coupled to each other. Raman scattering through a measurement of the vibrational frequency of the mode in question thus serves as a medium range and extended range probe of the elastic behavior of glasses particularly as the Ge concentration increases and the extended regions around each tetrahedral unit begins to overlap. The mode frequency squared measured relative to its threshold value, $\nu_{CS}^2 - \nu_C^2$, provides a measure of the optical elastic constant (C-C$_o$) of the backbone with the constant of proportionality representing the inverse of Ge reduced mass. With increasing x, the backbone progressively cross-links, and its stiffening is reflected in a blueshift of the vibrational mode frequency. In region II, the results of Fig.3b show a power-law, p = 0.99(3). In region III, results of Fig.4 yield a power-law p = 1.48 (2). In the analysis, we have proceeded by fitting these data sets using both log-log plots and separately polynomial fits to equation 2. By steadily changing the phase boundary values, $x_c(1)$, $x_c(2)$, from an initial estimate, both types of fits converge to give the same value of the power-law p as illustrated in Fig. 4.

In the present glasses, once the P (or Ge) concentration x exceeds 18%, our Raman scattering data show some $P_4Se_3$ molecules to decouple from the backbone (Fig.2c and Fig.2d), and in m-DSC experiments a satellite window (Fig.3a) in the $\Delta H_{nr}$ term is manifested. These features are associated with a mild NSPS of the ternary glasses. Of interest here is the fact that once the glass demixes, the power-law variation (Fig.3b) is interrupted. These data corroborate the view that the CS mode frequency is, indeed, a good probe of network connectivity as long as the network remains fully polymerized.

The compositional trends in the non-reversing enthalpy term, $\Delta H_{nr}(x)$, inferred from m-DSC experiments are compared to those of the Raman CS mode frequency in Fig.3. The correlation between the reversibility window and region II is quite striking. In the reversibility window glasses are in the IP. In Region III, the elastic power-law of p = 1.48 (2) is in excellent agreement with the value predicted for stressed-rigid networks in numerical experiments [18, 19]. There are currently no theoretical predictions for the elastic power-law in IPs. Raman results on binary ($Si_xSe_{1-x}$ glasses[5] $Ge_xSe_{1-x}$ glasses[8]) and ternary ($Ge_xAs_xSe_{1-2x}$[6]) glasses reveal, for elastic power-laws in the IP and Stressed-Rigid phases, values that are quite similar to the ones observed here in the $Ge_yP_ySe_{1-2y}$ ternary. A summary[22] of Intermediate Phases in several chalcogenide glasses appears in Fig. 5 as a bar chart, with the length of the bar delineating the IP-width in *r.*

Another significant finding of the m-DSC results (Fig. 3) on the present ternary glasses is the aging behavior of the $\Delta H_{nr}(x)$ term. In these experiments glass samples were aged at room temperature, $T < T_g$, and glass transitions studied for periods up to 6 months after synthesis. We find that the $\Delta H_{nr}(x)$ term for glass compositions in the flexible and stressed-rigid regimes ages, but such aging is qualitatively absent for IP glass compositions. This is a remarkable feature of IP glasses and has been discussed elsewhere[7].

Trends in IP- widths of several covalent glasses are summarized in Fig.5. We note that, in general, group V- selenides have IPs that lie below *r* < 2.40 while the group IV-selenides have IPs that lie at r > 2.40. Ternary glasses of both group IV- and V- selenides have IPs that are much wider, and generally encompass the previous two ranges characteristic of the group IV and V –selenides. On the other hand, IPs in chalcohalide are found to be extremely narrow[9, 52] in width probably because halogen atoms serve to scission chalcogen bridges and break characteristic rings where rigidity is nucleated.

What is so special about these Intermediate Phases? These networks are usually composed of local and medium range structures that are isostatically rigid, i.e., that are rigid but with no redundant bonds (that create stress). In ternary $Ge_xP_xSe_{1-2x}$ glasses, the isotatic local structures populated[7] in the IP include, CS $GeSe_4$ tetrahedra ($r$ = 2.40), Pyramidal $P(Se_{1/2})_3$ units ($r$ = 2.40) and Quasi Tetrahedrally coordinated $Se=P(Se_{1/2})_3$ units ($r$ = 2.28). For each of these units, the count of constraints due to bond-stretching and bond-bending forces per atom yields, $n_c$ = 3, even though their mean coordination numbers, $r$ (indicated in parenthesis) change as reflected by their chemical stoichiometries. Raman pressure experiments[22] on glasses have shown that networks in IPs are stress-free and that in the Flexible and the Stressed-Rigid elastic Phases have internal stresses. IP glasses exist in a state of quasi-equilibrium by virtue of an ideal connectivity at all length scales[21] that is largely driven by these systems lowering their free energy by expelling redundant bonds. The absence of aging in IP glasses is a manifestation of the quasi-equilibrium of network structure. Thus, even though IP glassy networks are disordered, in so far as network stress is concerned, they behave much like crystalline solids. The reversibility of the glass transition and its non-aging characteristic for IP glass compositions captures the basic principles that underlie self-organization of disordered networks.

**GIANT – PHOTOCONTRACTION (PC) OF OBLIQUELY DEPOSITED $Ge_xSe_{1-x}$ THIN-FILMS AND INTERMEDIATE PHASES.**

The binary $Ge_xSe_{1-x}$ glass system was one of the first in which existence of the IP was established[8]. It spans the $0.20 < x << 0.25$ range as shown in Fig.6. In the 1980s, K.L.Chopra and collaborators showed[53] that obliquely deposited $Ge_xSe_{1-x}$ thin-films are porous and possess a columnar structure. When such films (typical thickness t ~1μm) are irradiated by a Hg vapor lamp, they undergo a contraction with fractional changes in thickness ($\Delta t/t$) of as much as 25%. Thin- films of the Ge-As-Se ternary[54] showed a rather large PC effect of nearly 25% near a mean coordination of $r$ ~ 2.40. The growth of columns in obliquely deposited films is not unique to amorphous chalcogenides, but the giant PC of such films after exposure to super band gap light certainly is. The molecular origin of this giant PC effect has remained largely speculative[55] to date.

We have recently re-examined obliquely deposited thin-films of $Ge_xSe_{1-x}$ over a wide composition range x ( $0.15 < x < 0.33$). The films were synthesized in Professor Chopra's laboratory at IIT, New Delhi, using an experimental set up in which thin-films at 5 obliqueness angles ( $\alpha = 0, 20°, 45°, 60°$ and $80°$) were deposited simultaneously using one evaporation charge. Si wafers and glass slides were used as substrates, and film thicknesses in the 1-2 μm range were deposited. In all cases evaporation charges used were the corresponding bulk glasses that were characterized by Raman scattering. Films were examined in SEM, Raman scattering, IR reflectance, and profilometry measurements both in the virgin (as deposited state), and in the post-illuminated state. Films mounted on to a cold stage and purged with nitrogen gas were exposed to a Hg-Xe vapor lamp to appropriate integrated doses of visible radiation and PC observed. Fig.7 shows an SEM of a $GeSe_2$ film obliquely deposited at $\alpha = 80°$. The columnar structure is readily observed in the SEM viewgraph. The bottom panel shows a cartoon of the film microstructure, which consists of columns of a phase labelled A, and an inter-columnar phase labeled as B. Columns usually grow at an angle β with respect to film normal when the atomic vapor is directed at an angle of incidence α. Dirk and Leamy have shown[56] that the angles α and β are related as, $\tan \alpha = 2 \tan \beta$.

Fig 6 shows trends in photocontraction of $Ge_xSe_{1-x}$ films. It is seen to be largest in the IP. We have also done Raman scattering on films as a function of obliqueness angle α at several compositions. Fig 8 compares the lineshapes of $GeSe_2$ thin-films deposited at three obliqueness angles ( 0°, 60° and 80°) with that of the stoichiometric bulk glass used as the evaporation charge. Several features are seen in these results. The film deposited at normal incidence ($\alpha = 0°$) displays vibrational features that are quite similar to those of the bulk glass, thus serving as a test of the thin-film deposition process. Tuning evaporation parameters is crucial to obtain films possessing a molecular structure similar to that of the bulk glass. On the other hand, we find that films deposited at high obliqueness angles (of 60° and 80°) progressively phase separate into a Ge-rich and a Se-rich phase that can be approximately described by the following stoichiometric relation:

$$3 \text{ GeSe}_2 = \text{GeSe}_3 + \text{Ge}_2\text{Se}_3 \qquad (4)$$

Analysis of the Raman lineshapes[57] reveals that the stoichiometry of the Se-rich phase is close to $Ge_{23}Se_{77}$, while that of the Ge-rich phase close to $Ge_{42}Se_{58}$. Since the band-gap of the Ge-rich phase ( < 1.5 eV) is smaller than that of the Se-rich phase (2.0 eV), and particularly since the latter gap is slightly larger than the laser energy (1.9eV, 647 nm) used, Raman scattering from films is dominated by resonant enhancement of the Se-rich phase. The principal features of the Se-rich phase in obliquely deposited α = 80° films include (i) presence of $Se_n$ chain mode near 260 cm$^{-1}$, (ii) a red-shift of the CS mode, and (iii) a reduction in scattering strength ratio of the ES to CS mode in relation to the normally deposited film ( α= 0°) as shown in fig.8. These features uniquely fix the stoichiometry[57] of the Se-rich phase to be close to the centroid of the reversibility window (Fig. 6) . The principal feature of the Ge-rich phase is a steadily declining scattering strength of the mode near 180cm$^{-1}$ as α increases to 80°. We identify this with a normal mode of ethanelike $Ge_2Se_6$ units[58]. In sharp contrast, Raman scattering results on thin-films at x = 0.23 ( Fig. 9) are found to display lineshapes that are largely *independent* of obliqueness angle α. The observation suggests that the deposited phase appearing at this composition is the IP, which is rigid but stress-free can easily grow in either columns or as a bulk 3d network.

It appears that two basic phenomena come together for the PC effect to be manifested in the present films. First, the morphological structure of the obliquely deposited (α= 80°) thin-films consist of two phases; a phase A that undergoes facile photo-collapse and a second phase B that does not. Films at x = 0.23 consist entirely of the generic A phase which can exist either in a columnar form or as a 3d network of interlocking rings. Films at x > 0.25, such as stoichiometric $GeSe_2$, consist both of a phase A (columns) and an intercolumnar Ge-rich phase B, as revealed by Raman scattering (Fig.8). A parallel situation prevails in films at x < 0.20, which segregate into phase A (columns) and an intercolumnar Se-rich phase B. As mentioned earlier, many materials can be obliquely deposited to form columnar structures but only the amorphous thin-films of the chalcogenides undergo pronounced PC effects. The second crucial feature of chalcogenides glasses is that phase A undergoes facile photo-melting. Photomelting of bulk $Ge_{23}Se_{77}$ glasses has been observed in micro-Raman scattering experiments as discussed elsewhere[8]. Super-band-gap radiation leads to switching of covalent bonds by an electronic process that has been described elsewhere[59] by H. Fritzsche. The compositional trends in the PC effect illustrated in Fig. 7 follow naturally from these considerations. The PC- effect is maximized when only the A-phase is in the reversibility window, $0.20 < x < 0.25$, since it undergoes photomelting and leads to a collapse of the columns. At $x > 0.25$ the inter-columnar Ge-rich phase B grows at the expense of the phase A, and its growth steadily suppresses the PC effect because it is intrinsically stressed and is not expected to undergo photomelting. Parallel considerations apply to thin-films at $x < 0.20$, where growth of the Se-rich B-phase serves to suppress the PC effect (Fig. 6) largely because it is also stressed[22].

The segregation of bulk $Ge_xSe_{1-x}$ glasses into two phases A and B upon oblique deposition (equation 4 reaction going to the right) can be partially reversed when films are photo contracted upon exposure to Hg vapor lamp.. The chemical bonding or reconstruction of phase A with phase B is supported by Raman scattering. We have examined $GeSe_2$ thin-films after Hg-light exposure, and in Fig.10 compare the observed lineshape of a virgin film with its post-illuminated counterpart. The super band gap radiation (470 nm , 2.63 eV) from Hg-lamp penetrates only to a small depth in obliquely deposited films. Such is not the case with the 647 nm excitation used to excite Raman scattering that essentially probes the total film thickness. The reduction in scattering strength of the chain mode near 260 cm$^{-1}$ and the enhancement in scattering strength of the ES mode ( 216 cm$^{-1}$), both normalized to the CS mode (200 cm$^{-1}$) serve to illustrate that some part of the phase separation has been reversed by Hg-light exposure. In summary, it appears that the giant PC effect is a manifestation of photomelting of the self-organized columns formed in IP compositions of these obliquely deposited thin-films.

**INTERMEDIATE PHASES IN SOLID ELECTROLYTE GLASSES**

Alloys of base oxide ( $B_2O_3$, $AgPO_3$, $GeO_2$) or chalcogenide ( $As_2S_3$, $GeSe_4$) glasses with solid electrolytes (alkali oxides or AgI, $Ag_2S$ and $Ag_2Se$) are widely known as solid electrolyte glasses. They form homogeneous glasses displaying a single glass transitions, but exceptions occur. That result in itself is quite surprising although it has never been addressed satisfactorily in the literature. Solid electrolyte glasses have attracted much interest as sensors, as materials for battery applications and flat panel displays[60]. The electrical conductivity of these materials can be increased by several orders of magnitude by alloying increasing amounts of a solid electrolyte additive, but quantitative details are sparse[61]. Recently it was recognized[62] that alloys of AgI and $Ag_2Se$ can exist in a fast-ion conducting glassy phase. Their $T_g$s were established in m-DSC experiments and the results rationalized in terms of their global connectivity based on constraint counting algorithms. Recently Novita et al.[63] and separately C.Holbrook et al.[64] have observed reversibility windows in two prototypical oxide and chalcogenide based solid electrolyte glass systems respectively. Furthermore, they have found that sharp thresholds for ionic conduction coincide with the three elastic phases of glassy networks. Activation energies for ionic conduction, $\Delta E_A$ decrease sharply when glasses enter the IP from the stressed-rigid and, furthermore, when the rigid glasses become flexible. These results open a new paradigm in the study of electric transport in solid electrolyte glasses.

m-DSC results on dry $(AgPO_3)_{1-x}(AgI)_x$ glasses[65] show glass transition temperatures to steadily decrease as the AgI content of the glasses increases. This is in harmony with the low effective connectivity of AgI glass as discussed earlier[62]. Previous work in the field[61, 66] has also found such compositional trends in the present $AgPO_3$-AgI glasses except that the reported $T_{gs}$ are almost 50°C to 80°C lower than the ones found by D.Novita et al.[65] The difference can be traced to the presence of bonded water in these glasses that lowers the connectivity, and thus $T_g$ of the glasses. The central finding of the m-DSC is the observation of a broad reversibility window in the 9.5% < x < 37.8% AgI range, which, in analogy with chalcogenide glasses[4-9], fixes the three elastic phases : glasses in the 0 < x < 9.5% range are viewed to be stressed-rigid, those in the 9.5% < x < 37.8% to be in the IP while those at x > 38% to be in the flexible phase. The room temperature electrical conductivity of these glasses also has a threshold near x = 9.5%, and a second one near x = 37.8%, the elastic phase boundaries suggested by the thermal measurements.

Raman scattering again provides important clues on the molecular structure of these glasses. Figure 11 captures the observed lineshape of the base $AgPO_3$ glass. It is dominated by two modes, one near 684 cm$^{-1}$ and the other near 1140 cm$^{-1}$. The base $AgPO_3$ glass is widely recognized to be composed of chains of $PO_3$ units (or $Q^2$ units in NMR notation) having two bridging ($O_b$) and two terminal ($O_t$) Oxygen near neighbors of P. Here $Q^n$ represents the number of P-cations that have n bridging oxygen near-neighbors. The modes near 1140 cm$^{-1}$ (PO$_t$) and 684 cm$^{-1}$ (PO$_b$) are characteristic of polymeric chains made of PO$_3$ units. With increasing AgI content, one observes a systematic evolution of modes; chain mode strength steadily goes over into ring modes of both small and large size. Assignment of the ring modes is facilitated by the tetrameric form[67] of c-AgPO$_3$ and other reference crystalline phosphates where large rings are manifested[68]. The central message contained in the Raman results is that there is a morphological transition in glass structure from polymeric at low x ( < 25%) to monomeric in character as x > 30%. AgI additive serves to decouple the network of chains by steadily inserting itself between chains at low x, and converting long chains to closed rings at high x. These observations that are broadly consistent with the earlier MD simulations[69] and molar volume results[70]. In the MD simulations a conductivity percolation transition was predicted near x ~ 0.3 which is consistent with the conductivity increase observed in these glasses. However, the observation of an IP in these glasses suggests that mean-field descriptions[69] of these glasses are unlikely to correctly get the details of these compositional trends in conductivity.

Of special interest is the vibrational mode near 1140 cm$^{-1}$, which is associated with P-O$_t$ bonds in long chains. Through the two terminal oxygen near-neighbors in PO$_3$ this particular mode serves to probe the connectedness of the AgI modified AgPO$_3$ base glass network. As the P-O-P- chain spacings in the base glass (AgPO$_3$) *increase* upon AgI alloying, the frequency of the P-O$_t$ mode steadily red-shifts (Fig.12). It, therefore, plays a role that is analogous to that of the corner-sharing GeSe$_4$ mode in binary Ge-Se glasses. Since the mode is rather well resolved in the Raman spectra (Fig.12), one can analyze the lineshapes and extract the compositional variation of the mode frequency quite accurately. The mode frequency displays two distinct thresholds, one near x = 9.5% and another near x = 37.8%. Furthermore, the underlying optical elasticity power-laws in the 0 < x < 9.5% range, and in the 9.5% < x < 37.8% range can also be established, and their values are found to be characteristic of those of stressed-rigid and IP glasses found in the covalently bonded networks. These new findings on solid electrolytes will be presented in a forth coming publication. These results open a new paradigm to understand fast-ion transport in terms of the elastic response of their networks. For the purpose of this review, these findings serve to illustrate the usefulness of Raman scattering as an elegant probe of IPs in disordered networks through measurements of *optical elasticity power-laws.*

**CONCLUDING REMARKS**

The discovery of Intermediate Phases has opened new vistas in understanding the disordered state of condensed matter[4, 7 13, 21]. In this review we have illustrated applications of Raman scattering as a probe of elastic phases in glassy networks, and in particular in characterizing IPs by their optical elasticity power-laws. Three types of material systems were considered, a chalcogenide bulk glass, a chalcogenide amorphous thin-film and a solid electrolyte bulk glass. IP glassy networks are rather special because they form ideal, space filling, and stress-free networks, with unusual functionalities. Absence of aging, giant photocontraction in obliquely deposited chalcogenide films, and a pronounced increase in electrical conductivity represent some of the functionalities displayed by IP glasses in the three examples chosen here. Disordered glassy networks form over a very small part of composition phase space; and IP glasses select an even smaller part of that phase space where the glass forming tendency is optimized. This is in contrast with amorphous networks rendered disordered by virtue of synthesis, such as vapor deposition onto cold substrates that may or may not display a glass transition. Amorphous thin-films of IP glass compositions in the Ge-Se binary also appear to display unusual functionalities, they form a columnar structure that can undergo facile photocollapse accounting for the giant photocontraction observed in these films. Intermediate phases have now been identified[21, 24, 25] in numerical simulations on triangular networks in 2d, and their physical properties including the feature of self-organized criticality, are being explored further. As these investigations evolve, a better understanding of the structural features at a local and medium range scale that endow these systems with their unusual functionalities can be expected to emerge. Window glass is an example of a self-organized oxide glass network whose chemical composition can now be derived rather accurately from two basic principles of aggregation and structure[10]. The importance of these generic elastic phases of disordered networks also have significant consequences in Protein Science[12]. They are beginning to assist in understanding the pathways taken by proteins when they fold and unfold reversibly. Intermediate phase glasses most likely represent the ultimate self-organized nanostructured functional materials optimized by nature.

**Acknowledgements**

We have benefited from on going discussions with J.C. Phillips, Ping Chen, Bernard Goodman, Darl McDaniel and M.Micoulaut . This work is supported by NSF grant DMR-04 -56472.

| Mode labels | Mode Frequency (cm$^{-1}$) | Mode Assignment | Reference |
|---|---|---|---|
| **CS** | 198 | Ge(Se$_{1/2}$)$_4$ corner sharing tetrahedra | K. Jackson et. al.[43] |
| **ES** | 215 | Ge(Se$_{1/2}$)$_4$ edge sharing tetrahedra | K. Jackson et. al.[43] |
| **F2** | 115, 130, 308 | Ge(Se$_{1/2}$)$_4$ F2 mode | P.M. Bridenbaugh et. al.[44] |
| **CM, A$_1$** <br> **CM, E** | 237, 256 <br> 138 | Se$_n$ chain (A$_1$ mode) <br> Se$_n$ chain (E mode) | R. Zallen and G. Lucovsky[42] |
| **RM** | 112 | Se$_8$ rings (A$_1$ mode) | R. Zallen and G. Lucovsky[42] |
| **ET** | 175, 208 | Ge$_2$(Se$_{1/2}$)$_6$ ethane like units | K. Jackson et. al.[43] |
| **PYR** | 210, 330 | P(Se$_{1/2}$)$_3$ pyramids | D.G. Georgiev et. al.[45] |
| **ETH** | 180, 330, 350, 370 | P$_2$(Se$_{1/2}$)$_2$Se$_2$ ethylene-like units | D.G. Georgiev et. al.[45] |
| **QT** | 500 | Se=P(Se$_{1/2}$)$_3$ quasitetrahedral units | D.G. Georgiev et. al.[45] |
| **P$_4$** | 363, 465, 606 | P$_4$ clusters | C.S. Venkateswaran[49] <br> K. Andreas et. al.[50] |
| **P$_4$Se$_3^m$ A1** <br> **P$_4$Se$_3^m$ E** | 184, 212, 368, 371, 483 <br> 135, 165, 425 | P$_4$Se$_3$ monomers (A$_1$ mode) <br> P$_4$Se$_3$ monomers (E mode) | D.J. Verrall et. al.[46] <br> D.G. Georgiev et. al.[45] |
| **a-P** | 355, 389, 398, 425, 458, 500 | Amorphous P | J.S. Lannin et. al.[48] |
| **a-Ge** | 230 | Amorphous Ge | S. Choi et. al.[47] |

**Table 1.** Mode labels, mode frequency, mode assignment and references for the Raman vibrational modes observed in ternary Ge$_x$P$_x$Se$_{1-2x}$ glasses**.**

**Figure Captions**

**Figure 1.** Variations in $T_g(r)$ in binary $Ge_xSe_{1-x}$, $P_xSe_{1-x}$ and ternary $P_yGe_ySe_{1-2y}$ glasses. $r = 2(1+x)$ for the $Ge_xSe_{1-x}$ system, $r = 2+x$ for the $P_xSe_{1-x}$ system and $r = 2 + 3y$ for the $Ge_yP_ySe_{1-2y}$ ternary. Inset shows the expected variation in the count of homopolar bonds Ge-Ge and P-P in the ternary glasses based on variations in $T_g$ predicted by Stochastic Agglomeration Theory. Fig taken from ref. 7

**Figure 2.** Raman scattering in $P_xGe_xSe_{1-2x}$ glasses excited by 647 nm radiation and analyzed by a triple subtractive monochromater system ( model T64000 from Jobin Yvon Inc) using a CCD detector in a macro –chamber. See Table 1 for the mode assignments. SOS= Second order Scattering, QT: quasi-Tetrahedral, PYR: Pyramidal, CM: Se chain mode, RM:$Se_8$ ring mode.

**Figure 3.** Compositional trends in the non-reversing enthalpy, $\Delta H_{nr}(x)$ are in the top panel, while trends in Raman CS mode frequency are in the bottom panel. The thermal and optical experiments reveal elastic phase boundaries that more or less coincide for $P_xGe_xSe_{1-2x}$ glasses.

**Figure 4**. Power-law (top) and polynomial fits (bottom) to the CS mode frequency squared in the intermediate phase (left panels) and the stressed-rigid phase (right panels) to deduce the elasticity power-laws in ternary $Ge_xP_xSe_{1-2x}$ bulk glasses.

**Figure 5**. Observed Intermediate phase widths (length of the red line) in r-space for indicated binary and ternary chalcogenide glasses.

**Figure 6**. Compositional trends in light induced contraction of obliquely deposited ($\alpha = 80°$) $Ge_xSe_{1-x}$ thin-films ( ▼ present work), and the non-reversing enthalpy (●) $\Delta H_{nr}(x)$ in bulk $Ge_xSe_{1-x}$ glasses

**Figure 7**. Scanning Electron Micrograph of an obliquely deposited thin-film of $GeSe_2$ showing the columnar structure in the top panel. A schematic model of the film structure is shown in the bottom panel. The obliqueness angle α indicated the angle made by the incident atomic vapor with the thin-film normal, while the angle β gives the angle between the column axis and the film normal. A and B designate the column and inter-column molecular phases populated in the films. See text for details

**Figure 8.** Raman scattering from obliquely deposited $GeSe_2$ thin-films at three angles ($\alpha = 0°, 60°$ and $80°$) compared to that of a bulk glass. The scattering was excited using 647 nm radiation.

**Figure 9.** Raman scattering in obliquely deposited $Ge_{23}Se_{77}$ thin-films at three angles ($\alpha = 0°, 60°$ and $80°$) compared to that of a bulk glass. The scattering was excited using 647 nm radiation.

**Figure 10.** Raman scattering of an obliquely ($80°$) deposited $GeSe_2$ thin-film taken before) and after Hg-light irradiation for 3 hours.

**Figure 11**. Raman scattering of a dry $AgPO_3$ base glass (taken from ref. 63) displaying several vibrational modes whose assignments are indicated in the figure.

**Figure 12**. Raman spectra of $(AgPO_3)_{1-x}(AgI)_x$ glasses showing a red-shift of the symmetric $PO_t$ mode of long chains with increasing AgI content. The shift shows that the network steadily softens as the interchain spacings increase.